\documentclass{article}

\usepackage{spconf,amsmath,graphicx}
\usepackage{multirow}
\usepackage{verbatim}
\usepackage{tabto}
\usepackage{enumitem}
\usepackage{xcolor}
\setlist{leftmargin=2.5mm}

\usepackage[
  separate-uncertainty = true,
  multi-part-units = repeat
]{siunitx}

\title{A study on native American English speech recognition by Indian listeners with varying word familiarity level}
\name{Abhayjeet Singh$^1$, Achuth Rao MV$^1$,  Rakesh Vaideeswaran$^1$, Chiranjeevi Yarra$^2$, Prasanta Kumar Ghosh$^1$}
\address{$^1$Electrical Engineering, Indian Institute of Science (IISc), Bangalore-560012, India\\
$^2$Language Technologies Research Center (LTRC), IIIT Hyderabad, 500032, India}
\begin{document}

\maketitle

\begin{abstract}
  In this study, listeners of varied Indian nativities are asked to listen and recognize TIMIT utterances spoken by American speakers. We have three kinds of responses from each listener while they recognize an utterance: 1. Sentence difficulty ratings, 2. Speaker difficulty ratings, and 3. Transcription of the utterance. From these transcriptions, word error rate (WER) is calculated and used as a metric to evaluate the similarity between the recognized and the original sentences. The sentences selected in this study are categorized into three groups: Easy, Medium and Hard,  based on the frequency of occurrence of the words in them. We observe that the sentence, speaker difficulty ratings and the WERs increase from easy to hard categories of sentences. We also compare the human speech recognition performance with that using three automatic speech recognition (ASR) under following three combinations of acoustic model (AM) and language model (LM): ASR1) AM trained with recordings from speakers of Indian origin and LM built on TIMIT text, ASR2) AM using recordings from native American speakers and LM built on text from LIBRI speech corpus, and ASR3) AM using recordings from native American speakers and LM build on LIBRI speech and TIMIT text. We observe that HSR performance is similar to that of ASR1 whereas ASR3 achieves the best performance. Speaker nativity wise analysis shows that utterances from speakers of some nativity are more difficult to recognize by Indian listeners compared to few other nativities.  
\end{abstract}
\vspace{-0.6cm}
\section{Introduction}
\vspace{-0.2cm}
A non-native listener of English finds it challenging to recognize  speech from a native English speaker. The recognition performance depends on listener's experience or exposure to Engilsh \cite{warzybok2015much}. The recognition ability is crucial for understanding of various online content such as Massive Open Online Courses (MOOCs). This is because a large percentage of the MOOC instructors is native speakers of American English. Hence, quantifying the difficulty of recognizing speech from native American English speakers is relevant in the Indian context. 

There are several studies on the recognition of L2 speech by a non-native listener. Several aspects have been explored in the past, such as different combinations of speakers' and listeners' nativity, background noise/reverberation, the effect of phonetic/word context, and the listener's proficiency. Studies in the literature showed that the native listeners are better at recognizing the native speech compared to non-native speech in quiet or low noise condition. Le et al.\cite{lecumberri2006effect} showed that the English listeners are better than Spanish listeners in the perceptions of English intervocalic consonants.  Cooke et al.\cite{cooke2008foreign} showed that the English listeners are better than the Spanish listeners in recognizing English keywords in a sentence.  There are various studies which deal with the effect of additive noise during listening by non-native listeners\cite{borghini2018listening,meador2000factors,scharenborg2018effect}. A summary of various studies and their findings can be found in \cite{scharenborg2019listening}.  These studies evaluate the accuracy of word recognition at different signal to noise ratios (SNRs).
The stimuli can be a sentence, isolated words, digits etc. Authors report that the accuracy deteriorates with a decrease in SNR. The native listeners consistently perform better than the non-native listeners.  They also found that the 2-6dB SNR adjustment is sufficient for the non-native listeners to perform on par with the native listeners \cite{bradlow2007semantic,cooke2008foreign}. Some studies compare the listeners in different noise types such as white noise \cite{bradlow2002clear}, babble noise \cite{mayo1997age,brouwer2012linguistic}, pink noise \cite{meador2000factors}, speech-shaped-noise \cite{cooke2008foreign}, etc. Nabvelek et al. have performed a study on the effect of reverberation in identifying vowels \cite{nabvelek1988identification}, and consonants \cite{nabvelek1984perception}. These studies showed that the gap in accuracy between native and non-native listeners increases as the listening conditions become more challenging. Some studies evaluate the effect of semantic context on target word recognition \cite{aydelott2012sentence}. Some of these methods have studied the effect of listener's proficiency \cite{warzybok2015much}. These studies include a wide variety of target language and non-native listeners--- English, Dutch, Spanish,  French, Swedish, Mandarin, Korean, etc. Some of these  also include a few Hindi and Tamil listeners \cite{bradlow2002clear}. Even though the word recognition performance gap  between the native and non-native listeners is studied, the existing studies do not focus on listeners from Indian nativity.  The effect of the speaker's and the listener's dialects and/or native languages have also not been studied. 

\begin{figure}
\centering
\centerline{\includegraphics[trim = 10mm 10mm 8mm 10mm, clip, width=6cm,height=4.5cm]{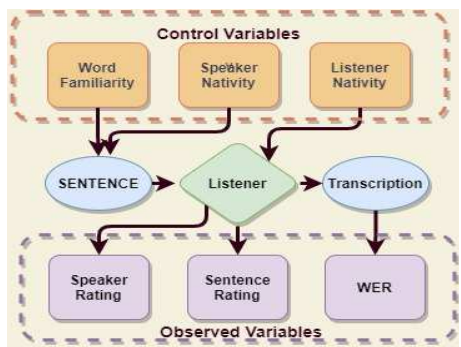}} %[width=10cm]
\caption{Overview of the study with different control and observed variables in the study. }
\label{blockdiagram} 
\vspace{-0.4cm}
\end{figure}

In this paper, we focus on the native American English  speakers and listeners from Indian nativity. We hypothesize that the difficulty of recognizing words in a sentence  depends on their frequency of occurrence (FoO-score)\cite{ANC}. We define a sentence's difficulty based on the sum of the FoO-score of the words in that sentence. We refer to it as word familiarity score (WF-score).  We hypothesize that the higher the WF-score of a sentence, the easier it is to recognize. Based on the WF-score, we categorize sentences into  three Word Familiarity levels (WFLs): easy, medium, and hard. 

The overview of different components of this study is shown in Fig. \ref{blockdiagram}. The study contains mainly three control variables: the word familiarity, the speaker's nativity, and the listener's nativity. In this study, an Indian listener is asked to provide two ratings to a sentence (1.Difficulty in understanding the sentence and 2. Difficulty in understanding speaker's accent) having a particular WF-score, spoken by a native American English speaker from the TIMIT corpus \cite{linguistic1990darpa}. These ratings reflect listener's self - assessment of the difficulty in recognizing the sentence and the difficulty in understanding the speaker. The listener also transcribes the sentence. We compute the word error rate (WER) \cite{spiccia2016semantic} for each transcription from a listener. Give these three observed variables (two ratings and one transcription) together with three control variables, we study the effect of different sets of control variables on these observed variables. We have found that the average speaker/sentence difficulty rating increases significantly when a sentence falls in the hard level compared to the easy level. However, it is found that Indian listeners find speech from native American speakers from Western dialect and Army Brat (moving around) speakers easier to understand and recognize compared to other dialect regions. The speaker and sentence difficulty ratings are found to be correlated to the WER. 

\vspace{-0.4cm}
\section{Materials used in study}
\vspace{-0.2cm}
For this study, 500 TIMIT \cite{linguistic1990darpa} sentences were selected and they were listened to \& recognized by  a total of 500 Indian listeners of varied nativities, where each listener recognized a subset of 10 sentences. TIMIT comprises 2342 unique sentences and, to select 500 sentences among those, we use American National Corpus (ANC) frequency dataset \cite{ANC}, which consists of nearly 250k unique words ordered in terms of their usage. We use this list of words to select and categorize the TIMIT sentences by following the algorithm.\\ \hrule
\vspace{2mm}
\textbf{Categorization Algorithm :}
\begin{enumerate} 
    \item[]1. For each sentence (Sent) in TIMIT corpus:
    \begin{enumerate}
        \item[]2. For each word (w) in Sent:
        \begin{enumerate}
            \item[]3. FoO-score(w) = index of w in ANC frquency dataset
            \item[]4. Sent.WF-score = sum(Sent.WF-score, FoO-score(w)); 
        \end{enumerate}
    \end{enumerate}
    \item[]5. Sort all 2342 sentences according to their score in ascending order;
    \item[]6. Divide the sorted sentence list into three lists of equal length;
    \item[]7. Select first 250, 150 and 100 sentences from the first, second and third list (referred to as Easy, Medium and Hard respectively) ;
\end{enumerate}
\hrule
\vspace{5mm}

With these 500 categorized TIMIT sentences we create 50 google forms using google apps scripts, each form containing 10 sentences: 5 from easy, 3 from medium and 2 from hard category. Each form contains unique set of sentences i.e., no sentence is repeated within a form as well as across all forms. Table \ref{speaker nativities} shows the speaker's dialect region (DR) distribution for the selected 500 TIMIT sentences over all the three levels: Easy, Medium and Hard. Sentence distribution across DRs is proportional to that in the TIMIT corpus. 500 sentences, thus selected, makes this study feasible while preserving the presence of each dialect in a manner similar to that in the TIMIT corpus.
The vocabulary of the selected 500 sentences comprises 1867 unique words, with 910, 556 and 401 unique words for easy, medium and hard level sentences, respectively. Two example sentences in each level is provided below in decreasing order of WF-score. Easy: 1) "Just let me die in peace", 2) "Leave me your address"; Medium: 1)"Such optimism was completely unjustified", 2) "Young children should avoid exposure to contagious diseases"; Hard: 1)"His captain was thin and haggard and his beautiful boots were worn and shabby", 2) "Eternity is no time for recriminations".

\begin{table}[h]
  \caption{Frequency of speaker Dialect Region (DR) in selected 500 TIMIT sentences.}
  \label{speaker nativities}
  \centering
  \vspace{2mm}
  \resizebox{0.48\textwidth}{!}{\large \begin{tabular}{p{4em} c c c c c c c c c}
    \hline
    \textbf{Dialect Region} & \textbf{DR1} & \textbf{DR2} & \textbf{DR3} & \textbf{DR4} & \textbf{DR5} & \textbf{DR6} & \textbf{DR7} & \textbf{DR8} & \textbf{Total} \\
    \hline
    \multirow{1}{4em}{\textbf{Easy}} & 68 & 95 & 28 & 17 & 14 & 9 & 14 & 5 & 250\\
    \multirow{1}{4em}{\textbf{Medium}} & 31 & 35 & 18 & 25 & 19 & 4 & 10 & 8 & 150\\
    \multirow{1}{4em}{\textbf{Hard}} & 18 & 26 & 9 & 19 & 10 & 8 & 8 & 2 & 100\\
    \multirow{1}{4em}{\textbf{Total}} & 117 & 156 & 55 & 61 & 43 & 21 & 32 & 15 & 500\\
    \hline
  \end{tabular}}
\end{table}

In each form, a respondent first listens to the selected TIMIT sentence audio clip after which a listener transcribes the sentence to the best of his/her ability. Following this, the respondent (listener) is shown the correct transcription of the sentence and is asked to  rate it on the basis of difficulty in recognizing the sentence as well as difficulty in understanding the speaker. 
The ratings are numbers ranging from 1 to 10, 1 representing 'Easy' and 10 representing 'Difficult'. The transcription provided by the listener is used to calculate the WER between the transcription and the original TIMIT sentence. WER is calculated using Kaldi\cite{povey2011kaldi}, which uses Levenshtein distance algorithm \cite{levenshtein1966binary}. This provides the minimum number of edits like insertion, deletion and substitution to convert one string to another.
\begin{table}[h]
\vspace{-0.6cm}
  \caption{Percentages of Indian listeners' nativities.}
  \label{listener nats}
  \centering
  \vspace{2mm}
  \resizebox{0.48\textwidth}{!}{\tiny \begin{tabular}{p{4em} c p{4em} c }
    \hline
    \textbf{Nativity} & \textbf{Percentage} & \textbf{Nativity} & \textbf{Percentage} \\
    \hline
    Hindi & 27.19 & Kashmiri & 3.23\\
    Telugu & 18.43 & Konkani & 1.84\\
    Tamil & 13.36 & Gujarati & 1.38\\
    Kannada & 11.98 & Urdu & 0.92\\
    Bengali & 10.60 & Odia & 0.92\\
    Malayalam & 5.07 & Others & 0.92\\
    Marathi & 4.18 & &\\
    \hline
  \end{tabular}}
  \vspace{-0.2cm}
\end{table}

This survey was conducted in an online mode, by circulating these forms in undergraduate colleges and universities across India. The objective was to get at least 10 responses from undergraduate students with ages ranging from 19 to 23 years with varied Indian nativities. Table \ref{listener nats} shows the distribution of listeners' nativities, where Hindi, Telugu, Tamil, Kannada and Bengali have major contributions (each above 10\%). For each form we received at least 10 responses from different listeners and only the first 10 responses to each form are taken into consideration while performing analysis in this study. With 50 forms and 10 unique responses to 10 sentences in each form, we have a total of 5000 responses, using which we carried out analysis as summarized in the next section.
\vspace{-0.3cm}

\section{Study Outcomes and Discussion}
\begin{figure}[h]
   \centering
   \centerline{\includegraphics[trim = 0mm 0mm 0mm 0mm, clip,width=\columnwidth]{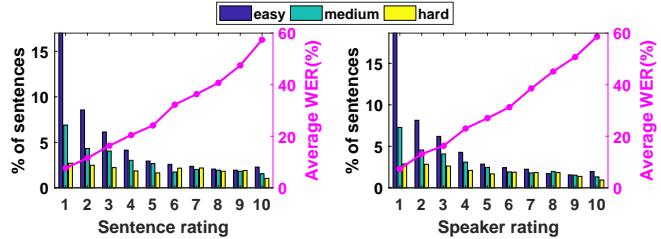}} %[width=10cm]
   	\caption{Percentage Word Error Rates across speaker and sentence difficulty ratings}
   	\label{wervsrating}
   	\vspace{-0.6cm}
 \end{figure}

\vspace{-0.2cm}
In this study, we investigate several aspects including the relation between the sentence difficulty and speaker accent difficulty ratings provided by the listeners as well as the relation between each of those ratings and the WER. We also study the variations depending on the dialect region of the native American speakers as well as native language of the listeners. We observe a correlation of 0.90 between the two ratings provided by the listeners. This implies that when a listener finds it difficult to follow a speaker, he/she finds it difficult to recognize the sentence as well. As both types of ratings are correlated, their relation with WER is also similar. As seen in the Figure \ref{wervsrating}, WER increases as the sentence difficulty rating goes higher. This is true for speaker difficulty rating as well suggesting that WER for Indian listener goes high not only with perceived difficulty to recognize a sentence but also with perceived difficulty to understand the speaker.  
\\ 
It can be observed from Figure \ref{wervsrating}, that the percentage of sentences corresponding to each sentence difficulty rating decreases as the ratings increase, which is expected as the ratio of sentences with easy, medium and hard WFL categories is 5:3:2. In each of the ratings from 1 to 10, histograms of percentage of sentences in each WFL are plotted. As we go from rating 1 to 10, there is a steady decrease in the percentage of easy and medium category sentences. However the percentages of hard sentences are more evenly distributed over the rating range. This indicates that the both ratings provided by the listeners closely follow the three WFLs (Easy, Medium and Hard) of the 500 sentences chosen for the study.
\vspace{-0.3cm}
\subsection{Effect of Word Familiarity}
\begin{figure}[h]
\vspace{-0.4cm}
\centering
\centerline{\includegraphics[trim = 0mm 0mm 0mm 0mm, clip, width=\columnwidth]{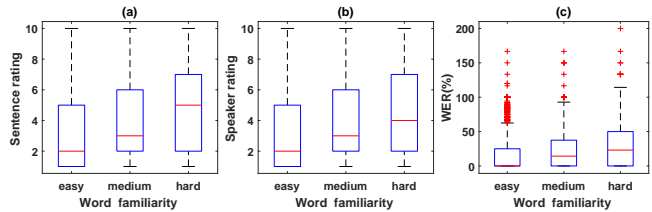}} %[width=10cm]
\vspace{-0.3cm}
\caption{a. Sentence difficulty rating increases with decreasing WFL i.e., easy to hard, b. Speaker difficulty rating increases with decreasing WFL, c. Human WER rises with decreasing WFL}
\label{familiarity}
\vspace{-0.4cm}
\end{figure}

\vspace{-0.1cm}
As shown in the Figure \ref{familiarity} (a,b), we analyze how the sentence and speaker difficulty ratings vary with WFL. We observe that the sentence difficulty ratings as well as the speaker difficulty ratings increase as WFL goes from easy to hard level. The average(SD) sentence difficulty ratings for easy, medium and hard categories are 3.25 ($\pm$ 2.62), 4.00 ($\pm$ 2.77) and 4.70 ($\pm$ 2.75) respectively  Similarly, average (SD) speaker difficulty rating for easy, medium and hard categories are 3.45 ($\pm$ 2.73), 4.16 ($\pm$ 2.85) and 5.01 ($\pm$ 2.79). All these ratings are significantly different from each other $(p<0.01)$. A significant increase of sentence and speaker difficulty ratings from high to low word familiarity, suggests that the word familiarity does play a significant role in the perceived difficulty in recognizing native American English speech by native Indian listeners.

In Figure \ref{familiarity}(c), we present the WER values calculated between transcripts provided by the listeners and the original TIMIT sentences for each WFL. We observe that WER (in percentage) value increases with decreasing WFL, i.e., from easy to hard . The average WER values for easy, medium and hard categories are 17.43\% ($\pm$ 25.6), 23.17\% ($\pm$27.01) and 30.78\% ($\pm$30.90) respectively. Percentages of zero WER for easy, medium and hard WFLs are 50.76\%, 36.8\% and 25.6\% respectively, this decline in percentages suggest that word familiarity significantly alters human speech recognition (HSR) performance.\\
\textbf{Key findings:} Word familiarity has a significant impact on the difficulty in recognizing (and, hence, recognition accuracy) a speech from native American speaker by listeners of Indian nativity. Word familiarity also makes it more difficult to understand a native American speaker.

\begin{table}[h]
\vspace{-0.5cm}
  \caption{Comparison of HSR vs ASR for different word familiarity level (WFL). [X]+\{Y\} indicate the AM trained using data ``X" and the LM trained using the data ``Y". }
  \label{hsr_vs_asr}
  \centering
  \vspace{2mm}
\resizebox{0.5\textwidth}{!}{\Huge \begin{tabular}{c c c c c}
\hline 
\multirow{2}{*}{WER(\%)} & \multirow{2}{*}{HSR} & 
ASR 1 & ASR 2 & ASR 3\tabularnewline
 &  & [iTIMIT]+\{iTIMIT\} & [LIBRI]+\{LIBRI\} & [LIBRI]+\{LIBRI+iTIMIT\}\tabularnewline
\hline
%\hline
\multirow{1}{*}{Easy} & 17.43 (25.6) & 29.70 (36.1) & 13.21 (18.4)  & 2.82 (10.8) \tabularnewline
%\hline 
\multirow{1}{*}{Medium} & 23.17 (27.0) & 21.05 (30.0)  & 20.67 (22.0) & 6.37 (21.1) \tabularnewline
%\hline 
\multirow{1}{*}{Hard} & 30.78 (30.9) & 28.47 (39.9) & 26.57 (30.5 ) & 10.02 (23.3)\tabularnewline
\hline
\end{tabular}}
\vspace{-0.8cm}
\end{table}

\subsection{ASR vs HSR for different word familiarity level}
\vspace{-0.2cm}
%\textcolor{red}{In this study native American listener's would have achieved the best performance but unfortunately we couldn't achieve this, so we have taken ASRs in their stead}. In order to understand how humans perform in the recognition task compared to an automatic recognition system trained with native American English as well as Indian English, we compare and analyze their performances.
As this study is performed on the native American speakers' recordings, native American listeners would achieve the best recognition performance. However, an ASR built on American English has been used as a proxy for a native American listener in this work. We also experimented with ASR built on Indian English data for a comparative analysis. In particular, three ASRs were trained 1) ASR1: Acoustic model (AM) for this ASR was trained using 81 Indian speakers speaking sentences from TIMIT corpus (iTIMIT $\sim$275 hrs)  \cite{9041230} and the language model (LM) was trained on the TIMIT data, 2) ASR2: Both AM and LM were trained using LIBRI speech corpus\cite{7178964} ($\sim$960 hrs), 3) ASR3: Same AM as of ASR2 but the LM was trained using both TIMIT and LIBRI speech text.

Table \ref{hsr_vs_asr} shows a comparison of WER for HSR and ASR for different WFL categories. It is clear from the table that the WER for both HSR and ASR decreases with increase in the WFL. This indicates that the ASR also performs poorly for the sentences with less familiar words.  This could be because of the lack of training data for these words. ASR1 performs poorly in case of easy sentences and performs better in case of hard/medium sentences. This could be because the AM performs poorly because of the accent mismatch and the LM helps more in case of difficult sentences. ASR2 performance is consistently higher than the HSR in all WFLs. This indicates that the AM trained with American accent corpus (LIBRI speech) further improves the WER. WER for ASR1 is comparable to the HSR. A significant  ($p<0.01$) decrease in WER is observed as we shift from AM trained on Indian accent data to LIBRI speech (native American accent). And in case of ASR3, where LM is trained on both LIBRI and TIMIT, performance of ASR improves significantly. This indicates that the inclusion of TIMIT data to LM significantly affects the ASR's performance  as LIBRI speech text covers a wide variety of English words and TIMIT text covers less familiar words as well which may have been missed in LIBRI speech.
\vspace{-0.6cm}
\subsection{Impact of Speaker Nativity on Speech Recognition}
\begin{figure}[h]
\vspace{-0.6cm}
\centering
\centerline{\includegraphics[trim = 0mm 0mm 0mm 0mm, clip, width=\columnwidth]{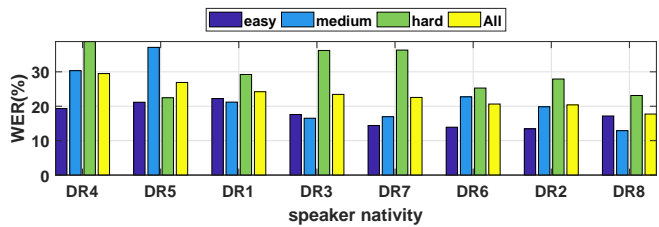}} %[width=10cm]
\vspace{-0.4cm}
\caption{Average WER calculated for Speakers across different dialects}
\label{wer_vs_drs}
\vspace{-0.4cm}
\end{figure}

\vspace{-0.2cm}
WER corresponding to speakers of all DRs (as summarized in Table \ref{speaker nativities}) is plotted in the Figure \ref{wer_vs_drs}. In this figure, we have average WERs for each DR in each WFL category (easy, medium, hard) and an overall. A trend in all DRs is observed that is, higher the familiarity lower the average WERs. DR8 and DR5 are exceptions to this observed trend. In Figure \ref{wer_vs_drs} the DRs are shown in descending order of the overall WER in all DRs. Average WER percentages for DR8, DR2, DR6, DR3, DR7, DR1, DR5 and DR4  are 15.69\% ($\pm$18.92), 17.54\% ($\pm$24.72), 19.92\% ($\pm$23.79), 20.29\% ($\pm$24.58), 20.68\% ($\pm$29.53), 23.04\% ($\pm$28.50), 28.51\% ($\pm$29.64) and 29.91\% ($\pm$31.03), respectively. It is observed that the WERs for the DRs at the extremities of the ordered DR list are significantly different $(p<0.01)$ from each other. For example DR8 and DR2 are significantly different from DR4, DR5 and DR1 whereas DR6 and DR3 are significantly different from DR4 and DR5. This suggests that dialect region of the native American speakers is a key factor to influence the recognition accuracy by Indian listeners.\\

\begin{figure}[h]
   \centering
   \centerline{\includegraphics[trim = 18mm 0mm 0mm 15mm, clip, width=\columnwidth, height=3cm]{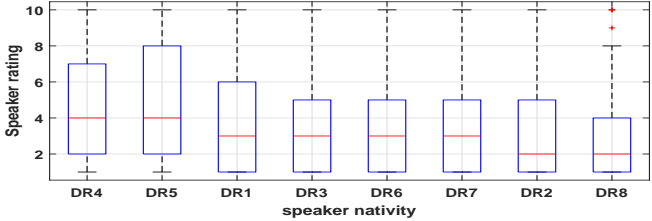}} %[width=10cm]
   	\caption{Speaker difficulty ratings vs different dialects}
   	\label{spkr_ratingvsnat}
   	\vspace{-0.5cm}
 \end{figure}

In Figure \ref{spkr_ratingvsnat} we present box-plots of speaker ratings for each DR. DRs in the plot are ordered following the average speaker difficulty rating.
Average speaker difficulty ratings for DR1 to DR8  are 4.41\% ($\pm$2.91), 3.86\% ($\pm$2.66), 4.14\% ($\pm$2.67), 4.72\% ($\pm$2.92), 4.96\% ($\pm$2.79), 3.99\% ($\pm$2.64), 3.72\% ($\pm$2.55) and 3.07\% ($\pm$2.25) respectively.
DR4 and DR5 have the highest WER numbers and are significantly different $(p<0.01)$ from rest of the nativities as shown in Figure \ref{wer_vs_drs} and this is also reflected in case of speaker difficulty ratings. This could indicate that Indian listeners' recognition accuracy on speech from American speakers vary across speaker's dialects and that may be influenced by how difficult it is to follow a speaker's speaking style that varies across different dialects. For example, with speaker difficulty rating in perspective, we find that DR1, DR4 and DR5 are significantly different $(p<0.01)$ from DR8, DR2 and DR7. Thus we can conclude that Indian listeners find speakers from DR8, DR2 and DR7 easier to follow significantly compared to DR1, DR4 and DR5. This could be a reason for a differences in WER between these two groups of dialect regions. 
\\ 

 \begin{comment}
textbf{Key findings:} Speaker nativity has a significantly impacted the HSR performance. As a definite clusterization of different dialect regions is seen such as cluster1 \{DR4, DR5, DR1\} and cluster2 \{DR8, DR2 and DR7\}. Indian speakers find speakers of cluster1 hard to follow as compared to cluster2. The speaker difficulty ratings and WER for cluster1 are significantly higher than cluster2. 
 \end{comment}
\vspace{-0.6cm}
 \subsection{Word Deviation and Familiarity}
 \vspace{-0.2cm}
 Apart from the WER, we also calculate the deviation of number of words between the transcript provided by the listener and the original sentence. This is done to examine if Indian listeners could identify the number of words, if not exact words, in a sentence spoken by native Americans. In Table \ref{word deviation} we have seven different deviation values between the number of words in the original TIMIT sentences and the transcription provided by the listener with zero implying an exact match in the number of words and ($<-2$) and ($>2$) refer to cases where original sentence has 2 less and 2 more words compared to those in the transcript, respectively. Negative word deviations in Table \ref{word deviation} means that the word in listener's transcript are more than that in the the original transcript. The negative deviations are less than their positive counterparts probably due to listeners missing the spoken words more compared to inserting new words. It is observed that only for exact matches (word deviation = 0) the percentage of deviations decrease as we go from easy to hard WFL categories, implying higher the familiarity, higher are the chances of exact match. And in rest of the cases the trend is reversed where the word deviations increases with drop in the familiarity level.
\vspace{-0.2cm}
\begin{table}[h]
\vspace{-0.2cm}
  \caption{Numbers in the table indicate the percentage of sentences with a particular deviation of the number of words in a WFL category.}
  \label{word deviation}
  \centering
  \vspace{2mm}
  \resizebox{0.47\textwidth}{!}{\begin{tabular}{p{4em} c c c c c c c}
    \hline
    \textbf{Deviation} & \textbf{0} & \textbf{1} & \textbf{-1} & \textbf{2} & \textbf{-2} & \textbf{$>2$} & \textbf{$<-2$} \\
    \hline
    \multirow{1}{4em}{\textbf{Easy}} & 75.16 & 10.8 & 4.72 & 2.88 & 0.8 & 5.48 & 0.16 \\
    \multirow{1}{4em}{\textbf{Medium}} & 67.87 & 10.6 & 8.93 & 3.87 & 1.07 & 7.27 & 0.4 \\
    \multirow{1}{4em}{\textbf{Hard}} & 58.3 & 13.6 & 13.2 & 3.4 & 2.1 & 8.8 & 0.6 \\
    \hline
  \end{tabular}}
  \vspace{-0.2cm}
\end{table}

Percentage of exact matches in the number of words between transcripts and original sentences are also calculated for all speaker nativities (DR1 to DR8): 66.84\%, 75.19\%, 72.55\%, 62.62\%, 61.63\%, 66.19\%, 71.25\% and 74.67\%, respectively. As discussed in section 3.3, Indian listeners found American speakers from DR4 and DR5 as the most difficult ones to understand. And this is reflected in the case of number of word deviations as DR4 and DR5 have the least percentages of sentences with exact matches in the number of words between the original sentence and the listener's transcription.
\begin{comment}
\textbf{Key findings:} Word deviation tends to be lower for sentences with high familiarity words. Implying that Indian listeners tend to recall sentences with high familiarity with high confidence but speaker's nativity also impacts the HSR performance. \\ 
\end{comment}
\\
We also perform grammar check on the listener's transcriptions using language-check library \cite{languagecheck}, which detects spelling mistakes, syntactical as well as semantic errors. Overall, 25.48\% of sentences have at least one type of grammatical error. 
It was found that grammatical error increases as the WFL is lowered, 19.04\% for easy, 29.47\% for medium and 35.6\% for hard. To check the extent of these grammatically incorrect sentences' effect on the WER, we calculated overall WER (21.83\%), WER for grammatically correct transcriptions (16.60\%) and for incorrect ones (30.95\%). We observe a significant ($p<0.01$) increase in WER as we go from correct to incorrect cases, suggesting a significant contribution of grammatically incorrect transcriptions in the overall WER.

\begin{table}[h]
\vspace{-0.5cm}
  \footnotesize
  \caption{Average Word Error Rate (WER) for most frequent nativities of the listeners. Standard deviation in brackets.}
  \label{nat}
  \centering
  \vspace{2mm}
  \begin{tabular}{p{4em} c c c c c c}
    \hline
    \textbf{} & \textbf{Telugu} & \textbf{Bengali} & \textbf{Kannada} & \textbf{Hindi} & \textbf{Tamil}   \\
    \hline
    \multirow{2}{4em}{\textbf{WER}} & 25.78 & 24.22 & 18.63 & 16.25 & 15.12  \\
    & (26.94) & (40.36) & (23.76) & (24.43) & (20.65)  \\
    \hline
  \end{tabular}
  \vspace{-0.3cm}
\end{table}

An analysis on the nativity of the Indian listeners and respective WERs is shown in Table \ref{nat}, where
we have the average WER for the five most prominent nativities. We find that Telugu which has the highest average WER is significantly different $(p<0.01)$ from Kannada, Hindi and Tamil whereas  Bengali which has second highest WER, is significantly different $(p<0.01)$ from Hindi and Tamil. This could be due to the amount of exposure listeners of these nativities had to American English, in addition to the nativity specific factors in recognition.
\vspace{-0.5cm}
\section{Conclusions}
\vspace{-0.3cm}
All the observed variables in this study such as sentence, speaker difficulty ratings and the WER were found to increase significantly as the familiarity of the words decreased. This implies a significant impact of WFL on the non-native listeners' perception of difficulty level and their speech recognition performance. Variations of observed variables were also analyzed with speaker's dialect region. Speakers from some regions (DR4 and DR5) were found to be difficult to be followed by Indian listeners. Listeners' nativity plays a significant role in the speech recognition performance. Performances of HSR and ASR trained on Indian AM and Indian LM were found to be similar, and if the Indian AM is replaced with American AM we observe a significant improvement.
Variations in the observed variables with the speech rate of the speakers was also analyzed but no significant trend was observed. The deviations in the number of words between original sentence and the transcription are seen to rise with lowering WFL, suggesting that Indian listeners find it difficult to to identify number of words in a sentence having words with low familiarity.
\vspace{-0.5cm}

%\section{Acknowledgements}
\bibliographystyle{IEEEtran}

\bibliography{mybib}

\end{document}